# Variational ansatz for the nonlinear Landau-Zener problem for cold atom association

A. Ishkhanyan[1], B. Joulakian[2], and K.-A. Suominen[3]

[1]*Institute for Physical Research NAS of Armenia, 0203 Ashtarak-2, Armenia*
[2]*LPMC, Université Paul Verlaine-Metz, 1 Bld Arago, 57078 Metz Cedex 3, France*
[3]*Department of Physics and Astronomy, University of Turku, 20014 Turun yliopisto, Finland*

**Abstract.** We present a rigorous analysis of the Landau-Zener linear-in-time term crossing problem for quadratic-nonlinear systems relevant to the coherent association of ultracold atoms in degenerate quantum gases. Our treatment is based on an exact third-order nonlinear differential equation for the molecular state probability. Applying a variational two-term ansatz, we construct a simple approximation that accurately describes the whole-time dynamics of coupled atom-molecular system for any set of involved parameters. Ensuring an absolute error less than $10^{-5}$ for the final transition probability, the resultant solution improves by several orders of magnitude the accuracy of the previous approximations by A. Ishkhanyan et al. developed separately for the weak coupling [J. Phys. A **38**, 3505 (2005)] and strong interaction [J. Phys. A **39**, 14887 (2006)] limits. In addition, the constructed approximation covers the whole moderate-coupling regime, providing for this intermediate regime the same accuracy as for the two mentioned limits. The obtained results reveal the remarkable observation that for the strong-coupling limit the resonance crossing is mostly governed by the nonlinearity, while the coherent atom-molecular oscillations arising soon after the resonance has been crossed are basically of linear nature. This observation is supposed to be of a general character due to the basic attributes of the resonance crossing processes in the nonlinear quantum systems of the discussed type of involved quadratic nonlinearity.

Since the realization of the Bose-Einstein condensates in dilute gases of neutral atoms [1, 2] the nonlinear version of the Landau-Zener term crossing problem [3, 4] has become a subject of considerable theoretical research [5-20]. Different nonlinear generalizations have been suggested and explored. Among these, of central interest is the basic case involving a quadratic nonlinearity in equations of motion due to its relevance to superchemistry [21], that is, coherent association of cold atoms into molecules via optical laser photoassociation [22] or magnetic Feshbach resonance [23]. In the context of cold atom association, the two-mode mean field Gross-Pitaevskii limit is described by the following semiclassical time-dependent nonlinear two-state model treating the atomic and molecular condensates as classical fields [21, 23, 24]:

$$i\frac{da_1}{dt} = U(t)e^{-i\delta(t)}\bar{a}_1 a_2,$$
$$i\frac{da_2}{dt} = \frac{U(t)}{2}e^{i\delta(t)}a_1 a_1,$$
(1)

where $a_1$ and $a_2$ are the probability amplitudes of atomic and molecular states ($\bar{a}_1$ denotes the complex conjugate of $a_1$), respectively, and the real functions $U(t)$, $\delta(t)$ describe the coupling between the two modes. In photoassociation, $U(t)$ is referred to as the Rabi frequency of the laser field, and the derivative $\delta_t(t)$ is the detuning of the laser field frequency from that of the transition from the atomic state to the molecular one. These functions are controlled by the applied optical (photoassociation) or magnetic (Feshbach resonance) fields. The Landau-Zener term crossing problem is now defined as a linear-in-time resonance crossing of the detuning, $\delta_t(t) = 2\delta_0 t$, the Rabi frequency being constant during the interaction, $U(t) = U_0 = $ const [3, 4].

We start our discussion by changing from system (1) to the equation for the molecular state probability $p = |a_2|^2$ written in the following form [11, 12, 25]:

$$\left(\frac{d}{dt} - \frac{1}{t}\right)\left[p'' - \frac{\lambda}{2}(1 - 8p + 12p^2)\right] + 4t^2 p' = 0, \tag{2}$$

where prime denotes differentiation with respect to time. Here, all the quantities are supposed to be dimensionless and we have introduced the conventional Landau-Zener parameter $\lambda = U_0^2/\delta_0$. System (1) describes a lossless process, where the total number of particles is conserved: $|a_1|^2 + 2|a_2|^2 = $ const $= 1$. Note that this normalization is incorporated in Eq. (2). Finally, we assume the initial condition of a pure atomic condensate, with no molecules available originally: $p(-\infty) = 0$.

Based upon our previous experience in the treatment of Eq. (2) (see, e.g., [26, 27, 28]) we introduce the following *two-term ansatz* involving three variational constants $A$, $C_1$, and $\lambda_1$:

$$p = p_0(A, t) + C_1 \frac{p_{LZ}(\lambda_1, t)}{p_{LZ}(\lambda_1, \infty)}. \tag{3}$$

Here, $p_{LZ}(\lambda_1, t)$ is the solution of the *linear* Landau-Zener problem for an effective $\lambda_1$ [26]:

$$\left(\frac{d}{dt} - \frac{1}{t}\right)\left(p''_{LZ} + 4\lambda_1 p_{LZ} - 2\lambda_1\right) + 4t^2 p'_{LZ} = 0, \tag{4}$$

and $p_0(A, t)$ is the solution of a nonlinear *augmented limit equation* controlled by an adjustable parameter $A$ [27,28]:

$$\left(\frac{d}{dt} - \frac{1}{t}\right)\left[-\frac{\lambda}{2}(1 - 8p + 12p^2) + A\right] + 4t^2 p' = 0. \tag{5}$$

Both $p_{LZ}(\lambda_1,t)$ and $p_0(A,t)$ are supposed to satisfy the initial condition $p(-\infty)=0$.

The linear Landau-Zener function $p_{LZ}(\lambda_1,t)$ is written in terms of known mathematical functions. For instance, it can conveniently be written in terms of the Kummer hypergeometric functions [29] (see, e.g., [11]). The solution produces the Landau-Zener exponential law for the final transition probability: $p_{LZ}(t=+\infty)=1-e^{-\pi\lambda_1}$. Note that the transition probability at the resonance crossing point $t=0$ also obeys an exponential dependence: $p_{LZ}(t=0)=(1-e^{-\pi\lambda_1/2})/2$.

Regarding the limit solution $p_0(A,t)$, integration of Eq. (5) via transformation of the independent variable, followed by permutation of dependent and independent variables, results in a *quartic* polynomial equation for $p_0$:

$$\frac{\lambda}{4t^2}=\frac{C_0+p_0(p_0-\beta_1)(p_0-\beta_2)}{9(p_0-\alpha_1)^2(p_0-\alpha_2)^2}, \quad (6)$$

where $C_0$ is an integration constant and the involved parameters $\alpha_{1,2}$, $\beta_{1,2}$ are defined as

$$\alpha_{1,2}=\frac{1}{3}\mp\frac{1}{6}\sqrt{1+\frac{6A}{\lambda}}, \quad \beta_{1,2}=\frac{1}{2}\mp\sqrt{\frac{A}{2\lambda}}. \quad (7)$$

For the initial condition $p_0(-\infty)=0$ it holds that $C_0=0$. Note that if now we take $A=0$, Eq. (6) is degenerated to a quadratic equation because three of the four parameters $\alpha_{1,2}$, $\beta_{1,2}$ become equal, $\alpha_2=\beta_1=\beta_2=1/2$. The solution to this quadratic equation diverges at $t\to+\infty$. Hence, it cannot be used as an appropriate initial approximation. In contrast, for a positive $\lambda/2>A>0$ the solution to Eq. (6) defines a bounded, monotonically increasing function which tends to a positive finite value less than $1/2$ when $t\to+\infty$ (Fig.1). This solution possesses all the necessary characteristics and, therefore, can be used as an appropriate initial approximation to construct an accurate solution to the problem. The introduction of the parameter $A$ is therefore a constructive step.

Though Eq. (6) does not determine $p_0$ explicitly, many important characteristics of $p_0(t)$ can be determined exactly. This includes the value of the function and its derivatives at the resonance crossing point $t=0$, as well as at $t\to+\infty$. For instance, the final value $p_0(+\infty)$ is easily found by noting that the left-hand side of Eq. (6) goes to zero as $t\to+\infty$. It is then seen that it should be $p_0(+\infty)=0$, $\beta_1$ or $\beta_2$. Since $p_0(t)$ is a monotonically increasing function with $p_0(-\infty)=0$, and since $\beta_2>1/2$, we deduce that $p_0(+\infty)=\beta_1$.

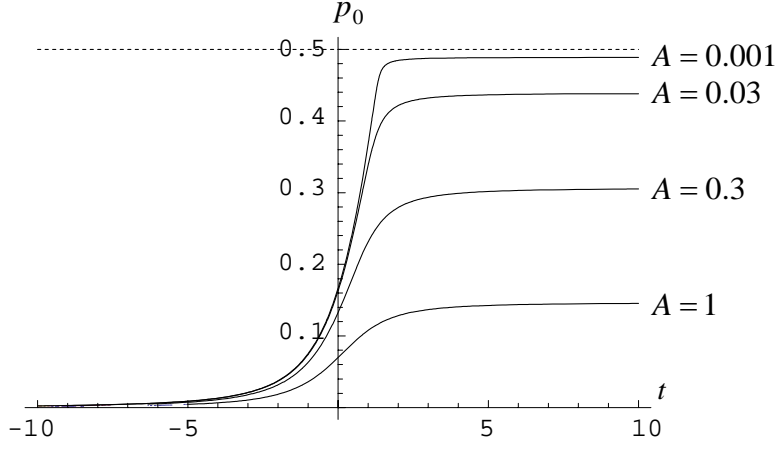

Fig.1. The limit solution $p_0(t)$ for a positive $A$ ($\lambda/2 > A > 0$) and fixed $\lambda = 4$.

In a similar way we find that $p_0(0) = \alpha_1$. Thus,

$$p_0(t=0) = \frac{1}{3} - \frac{1}{6}\sqrt{1 + \frac{6A}{\lambda}}, \quad p_0(t=+\infty) = \frac{1}{2} - \sqrt{\frac{A}{2\lambda}}. \tag{8}$$

Having introduced the ansatz (3), we first demonstrate numerically that it produces highly accurate results. Numerical simulations show that for any given value of the input Landau-Zener parameter, $\lambda \in [0, \infty)$, one can always find $A$, $C_1$, and $\lambda_1$ so that function (3) accurately fits the numerical solution to the exact equation for the molecular state probability (2) in the whole time domain – the graphs produced by the formula are practically indistinguishable from the numerical solution to Eq. (2) (see Fig. 2). More precisely, in quantitative terms, for any given $\lambda$, the proposed approximation assures an absolute error of less than $10^{-5}$ for the final transition probability $p(+\infty)$. For arbitrary time points, the absolute error is commonly of the order of $10^{-4}$ (the typical error curves for $\lambda \leq 1$ and $\lambda \gg 1$ are shown in Figs. 3a and 3b). The less accurate result is observed for points in a relatively small region embracing the first local maximums and minimums of $p(t)$ after the resonance crossing point has been passed: for this region, the error increases up to $\sim 10^{-3}$. Summarizing the observations above, we may state that the introduced ansatz describes the molecule formation process with very high accuracy in the whole time domain.

In Figs. 4, 5, and 6 we show the dependences $\lambda_1(\lambda)$, $A(\lambda)$, and $C_1(\lambda)$, respectively, obtained from numerical simulations (filled circles). These graphs suggest several general conclusions.

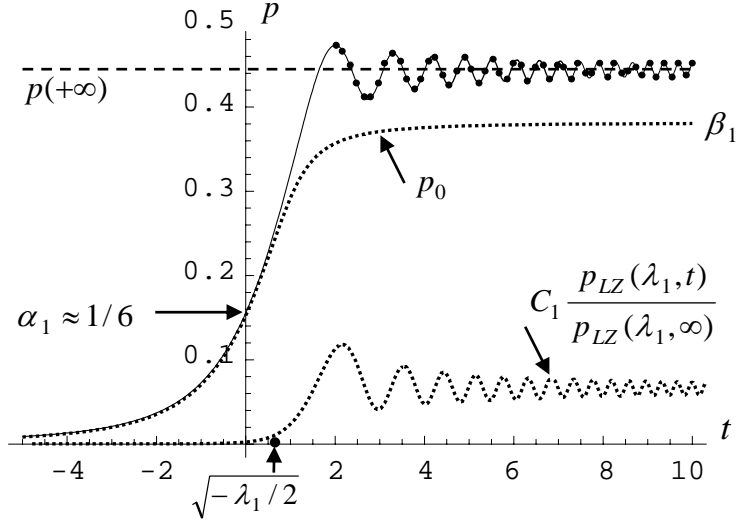

Fig. 2. Molecular state probability as a function of time at $\lambda = 4$ (dashed line is the final transition probability $p(+\infty) = \beta_1 + C_1$, dotted lines are the limit solution $p_0$ and the term proportional to $p_{LZ}(\lambda_1, t)$). The graph produced by formula (3) is indistinguishable from the numerical solution of Eq. (2). The filled circles are the time points used to fit formula (3). It is seen that in the strong coupling limit $\lambda \gg 1$ the prehistory of the system and the evolution near the resonance crossing region $t \approx 0$ are basically defined by the limit solution $p_0$, while the atom-molecule oscillations are described by the linear Landau-Zener solution with the effective Landau-Zener parameter $\lambda_1$.

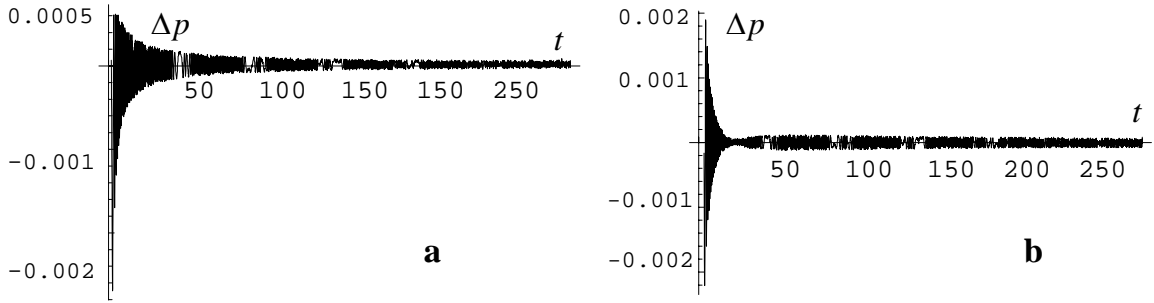

Fig. 3. Deviation of the approximation defined by formula (3) from the numerical solution of Eq. (2) at a) $\lambda = 0.7$ and b) $\lambda = 4$.

First, it is seen from Fig. 4 that for $\lambda \gg 1$, $\lambda_1$ is a large negative parameter. Apart from this unexpected sign, this observation leads to a more important conclusion. For a negative $\lambda_1$, the linear Landau-Zener function $p_{LZ}(\lambda_1, t)$ noticeably differs from zero not starting from a negative time interval preceding the resonance crossing at $t = 0$ (as it is the case for a positive Landau-Zener parameter), but exclusively for positive time points of the order of or larger than $t \approx \sqrt{-\lambda_1/2} > 0$ (see Fig. 2).

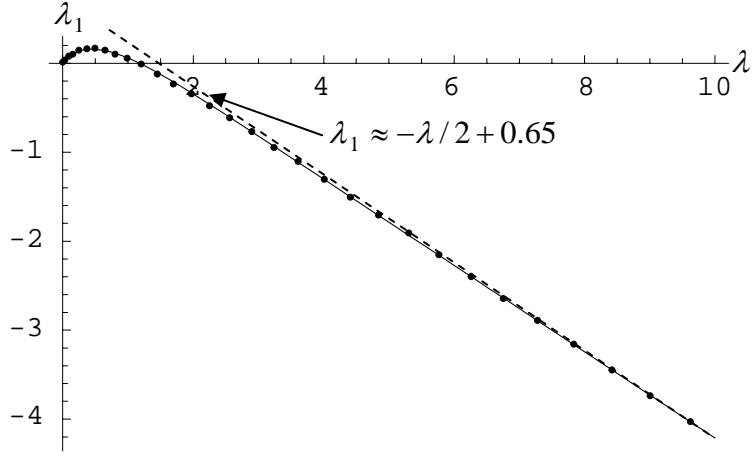

Fig. 4. Variational parameter $\lambda_1$ as a function of $\lambda$. Circles present the numerical fit for ansatz (3), the solid line presents the analytical formula (13). The two, weak and strong, interaction regimes differ in sign of $\lambda_1$: $\lambda_1$ is positive for $\lambda < 1$ while it becomes negative starting from $\lambda \approx \sqrt{\sqrt{2}}$. The asymptote of $\lambda_1$ for large $\lambda \gg 1$ is a linear function: $\lambda_1 \approx -\lambda/2 + 0.65$.

Hence, in the strong interaction limit of high field intensities $\lambda \gg 1$, the second term in the ansatz Eq. (3) is small when compared with the limit solution $p_0$, and thus can be effectively disregarded for the prehistory under $t < 0$ and a time interval after the resonance has been crossed. On the other hand, it is clearly seen from Fig. 2 that $p_0$ practically becomes constant at the end of the interaction, after coherent oscillations between the atomic and molecular populations have begun. Thus, in this final stage of the evolution the time dynamics of the system is basically controlled by the scaled linear Landau-Zener function $p_{LZ}(\lambda_1, t)$. In other words, since the limit solution $p_0$ is principally defined by the nonlinearity involved (see Eq. (5)), in the case of strong coupling the resonance crossing is mostly governed by the nonlinearity, while the coherent atom-molecular oscillations arising soon after the resonance has been crossed are basically of a linear nature. This decomposition is quite surprising as the equations of motion (1) do not indicate this.

Furthermore, a glance at the graphs of $A(\lambda)$ and $C_1(\lambda)$ (Figs. 5 and 6) immediately suggests that there exist two clearly discernible regimes of interaction: for $\lambda < 1$ we observe fast growth for the two parameters, while for $\lambda > 1$ we see a relatively slow decrease. The difference between the two regimes is also clearly seen in the behavior of the effective Landau-Zener parameter $\lambda_1(\lambda)$ (Fig. 4). Indeed, the two interaction regimes clearly differ in the sign of $\lambda_1$: $\lambda_1$ is positive for $\lambda < 1$, while it becomes negative starting from $\lambda \approx 1.17$. Further examination shows that in the limit of weak coupling (or, equivalently, fast sweeping), when $\lambda \to 0$, the parameter $A$ behaves as $A \sim \lambda/2$, while in the opposite limit of strong interaction (or slow sweeping), $\lambda \to \infty$, inverse dependence $A \sim 1/\lambda$ is observed.

The asymptotic behavior of the system in the limit of weak coupling is readily understood when examining the structure of ansatz (3), together with the properties of the limit function $p_0$. Indeed, it is clear physically and it also follows from Eq. (2) for the molecular state probability that in this limit the influence of nonlinearity ought to disappear. Since nonlinearity is manifested in Eq. (3) through the limit solution $p_0$, one should expect that $p_0(+\infty) \to 0$ when $\lambda \to 0$. This implies $\beta_1 \to 0$, which immediately leads to the asymptotes [26]

$$A \sim \lambda/2, \quad C_1 \sim P_{LZ}(\lambda, +\infty)/4 \text{ and } \lambda_1 \sim \lambda \text{ at } \lambda \to 0. \tag{9}$$

These are, indeed, the asymptotes observed from numerical simulations shown in Figs. 4, 5, 6. Furthermore, it has previously been shown [27] that in the opposite limit of strong interaction $\lambda \gg 1$ (high field intensities or slow sweeping rates) the asymptotic behavior of the system is

$$A \sim 1/\lambda, \quad C_1 \sim 1/\lambda \text{ and } \lambda_1 \sim -\lambda/2 + \text{const}, \quad \lambda \to +\infty, \tag{10}$$

where the constant is of the order of unity.

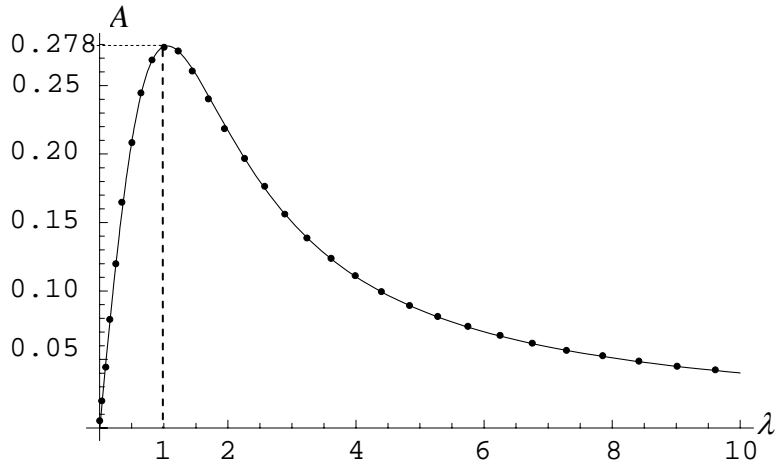

Fig. 5. Variational parameter $A$ as a function of the input LZ parameter $\lambda$. Circles present the result of the numerical fit using the ansatz (3) while the solid line presents the analytic formula for $A(\lambda)$ given by Eq. (11). Two clearly marked regimes of interaction are observed: the weak coupling regime corresponds to $\lambda < 1$, and the strong interaction occurs at $\lambda > 1$.

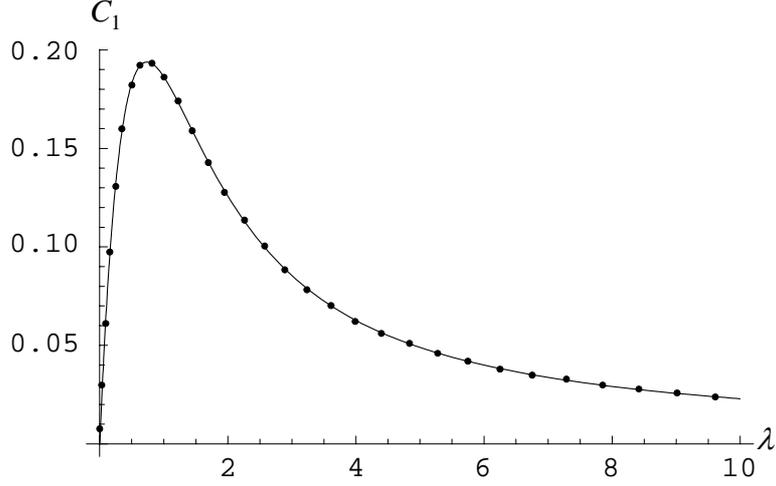

Fig. 6. Variational parameter $C_1$ as a function of $\lambda$. Circles present the result of the numerical fit for ansatz (3), the solid line presents analytic formula Eq. (12). Weak and strong interaction regimes correspond to increasing and decreasing branches of $C_1$, respectively.

We go further and find analytic approximations for the variational parameters $A(\lambda)$, $C_1(\lambda)$, and $\lambda_1(\lambda)$ that fit the numerical results much better than the previous asymptotes. To do this, we substitute the trial function (3) into the exact equation for the molecular state probability (2) and examine the remainder, more precisely, we consider in detail how the remainder will form the next approximation term. The minimization of the latter eventually leads to the formulas:

$$A = \frac{\lambda}{2} {}_2F_1\left(1,2;1.385;-\frac{\lambda^2}{2}\right), \qquad (11)$$

$$C_1 = \frac{P_{LZ}(\lambda,+\infty)}{4}\sqrt{{}_2F_1\left(1,2;1.284;-\frac{\lambda^2}{2.75}\right)} \qquad (12)$$

(${}_2F_1$ is the Gauss hypergeometric function [29]), and

$$\lambda_1 = \lambda\,(1 - 3\beta_1 - 3C_1) = \lambda\,(1 - 3p(+\infty)). \qquad (13)$$

The derived formulas define a fairly good approximation. Comparison of these formulas with the numerically found values of the introduced variational parameters is made in Figs. 4, 5, and 6. It is seen that the coincidence is, indeed, good. For the whole variation range of the input Landau-Zener parameter $\lambda$, deviation of the formulas from the numerical result does not exceed $10^{-4}$. Of course this ensures the same accuracy for the final transition probability $p(+\infty) = \beta_1 + C_1$, shown in Fig. 7. This result improves the previous approximations [7, 11, 12] by two orders of magnitude.

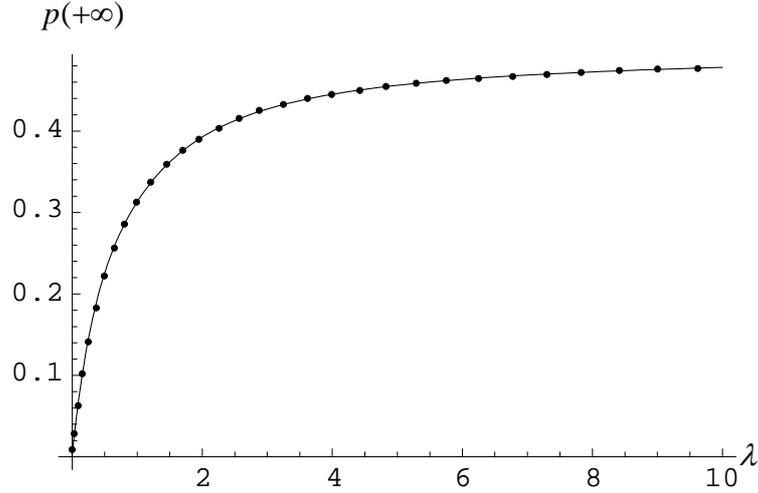

Fig. 7. Final transition probability to the molecular state as a function of $\lambda$. Circles show the result of numerical fit with ansatz (3), the solid line is calculated using formulas (11, 12), and we take
$$p(+\infty) = \beta_1 + C_1.$$

To summarize, we have introduced a two-term variational ansatz for the nonlinear Landau-Zener problem for coherent association of ultracold atoms. We have demonstrated numerically that this ansatz accurately describes the dynamics of the system in the whole time domain for any set of input parameters of the problem. It provides the final transition probability with an absolute error of less than $10^{-5}$, and for arbitrary time points the absolute error is mostly of the order of $10^{-4}$, increasing up to $\sim 10^{-3}$ in a relatively small region that includes the first local maximums and minimums of the transition probability $p(t)$ after the resonance has been crossed. The introduced ansatz involves three variational parameters, one of which serves as an effective Landau-Zener parameter in the linear Landau-Zener function involved in the proposed approximation. Surprisingly, this parameter proves to be a negative number in the strong interaction limit. The first term of the ansatz accounts for the nonlinearity, while the second one is basically of a linear nature. This decomposition leads to the conclusion that in the strong interaction limit (corresponding to high laser field intensities or slow sweep rates), the time evolution of the system can be divided into two different regimes: the prehistory and the very resonance crossing domain are mostly controlled by the nonlinearity, while the coherent oscillations between the atomic and molecular populations that begin after the resonance has been crossed are basically of a linear nature. This conclusion is applicable to all the level crossing models since it rests exclusively on the type of the quadratic nonlinearity discussed. Further, we have examined the asymptotes of the two parameters involved in the two aforementioned terms, and have shown that in the strong interaction limit they are inversely proportional to the input Landau-Zener parameter. Previously it was proposed that in the strong coupling limit the final transition probability to the molecular state obeys a power law [9, 10, 12-19]. The developed ansatz clearly shows that this is not strictly the case, because of the Landau-Zener exponential involved in the formula for $C_1$; rather, the power law is a good approximation if the accuracy of the description is not required to be very rigorous. We have proposed highly accurate approximate analytic formulas for the three involved variational parameters used. The expression for the final transition probability resulting from these formulas improves the previous result by an order of magnitude. Finally, we note that the proposed ansatz may be extended to other models, and it may be adapted to treat the extended versions of nonlinear two-state state problems involving higher-order nonlinearities, e.g., those representing the inter-particle elastic scattering. We have checked that this is the case for several physical situations (see e.g. [28]).

## Acknowledgments

This work was supported by the Armenian National Science and Education Fund (ANSEF Grant No. 2009-PS-1692) and the International Science and Technology Center (ISTC Grant N. A-1241). A. Ishkhanyan acknowledges Laboratoire de Physique Moléculaire et des Collisions, Université Paul Verlaine-Metz, France where this work was accomplished.